# Prediction of shock structure using the bimodal distribution function

Maxim A. Solovchuk,[1] and Tony W. H. Sheu[1,2*]

[1]Department of Engineering Science and Ocean Engineering, National Taiwan University, No. 1, Sec. 4, Roosevelt Road, Taipei, Taiwan 10617, Republic of China

[2]Center for Quantum Science and Engineering (CQSE), National Taiwan University

A modification of Mott-Smith method for predicting the one-dimensional shock wave solution is presented. Mott-Smith distribution function is used to construct the system of moment equations to study the steady-state structure of shock wave in a gas of Maxwell molecules and in argon. The predicted shock solutions using the newly proposed formalism are compared with the experimental data, direct-simulation Monte Carlo (DSMC) solution and the solutions predicted by other existing theories for Mach numbers M<11. The density, temperature, heat flux profiles and shock thickness calculated at different Mach numbers have been shown to have good agreement with the experimental and DSMC solutions. In addition, the predicted shock thickness is in good agreement with the DSMC simulation result at low Mach numbers.

47.40.Nm, 47.45.-n

---

Corresponding author. twhsheu@ntu.edu.tw

# INTRODUCTION

A normal shock wave is an example of highly non-equilibrium flows. An important parameter describing the non-equilibrium properties of the gas is Knudsen number, which can be defined in a shock wave as a relation between the mean free path and shock thickness. In the shock wave macroscopic properties of the gas change very fast within a short distance, which is about several mean free paths and the Knudsen number becomes large. The shock wave structure can not be described well by fluid dynamic equations in the sense that Navier-Stokes equations[1] give good agreement with the experimental data[2] [3,4] only at Mach numbers $M < 1.3$. Quite recently, normal shock wave has been studied using the Brenner's modification to Navier-Stokes equations (Brenner-Navier-Stokes).[5] Their results have a better agreement with the experimental data and Monte-Carlo simulations. When applying the Burnett and super Burnett equations some non-physical oscillations were found to appear in the solution even at M=2.[6] In Grad method[7] and extended irreversible thermodynamics,[8] a large number of equations must be solved to get a reasonable accuracy.[9] Grad's 13-moment method succeeded to simulate shock profile below the critical value $M_C = 1.65$. In Ref.[10] it was mentioned that one needs up to 680 moments (64 one-dimensional equations) to calculate a smooth shock structure for $M = 1.8$ that fits well to the experimental data. With the increasing number of moments in extended thermodynamics,[8] the solution converges rather slow. Therefore, a large number of moments is required to describe the processes at large Knudsen numbers. Good agreement with the experimental measurements was obtained on the basis of bimodal distribution function.[11] Mott-Smith pointed out that the distribution function in a strong shock wave is bimodal[11] and can be expressed by $f = a(x)f_0 + (1-a(x))f_1$, where

$f_0$ and $f_1$ are the local-equilibrium distribution functions for describing the supersonic and subsonic flows and $a(x)$ is an unknown quantity. Most of the experimental investigations of a gas or a plasma shock wave are devoted to the measurement of macroscopic parameters.[12] Only few works[12,13] have been directly devoted to the study of the distribution function across the shock wave. Those experimental work[12,13] as well as the recent molecular dynamic[14,15] and direct Monte-Carlo simulations[15,16] confirmed the main conclusions[17] about a bimodal distribution function in a shock region. The bimodal approximation of Mott-Smith may be considered as one of the most successful attempts to determine the structure of a planar shock wave by solving the Boltzmann equation.[14,15,18]

Because of its simplicity and correct prediction of shock thickness at large Mach numbers it was applied to several shock formation problems, including the shock structure in dense gases[11,16] and gas mixtures,[19,20] relativistic shocks,[21,22] plasma problem.[17] However, there still exist several nontrivial deficiencies in this theory[16,23,24,25,26]. The first drawback is that there is no unique way that is currently available to determine the unknown quantity $a(x)$, which needs to be determined from a moment equation given by the Boltzmann equation. The choice of velocity moment, while it can be arbitrary, can greatly affect the predicted result in the sense that the computed shock thickness can be different by an amount of 25%.[11,24] Bashkirov and Orlov[27] used non-analytical moments in velocity space and their results can have a difference about 80-100%. As a result, a better procedure should be adopted. The second deficiency is attributed to the incorrect prediction of shock thickness at low Mach numbers.[25] Our attempt in this paper is to get rid of these disadvantages. There are several approaches to improve the Mott-Smith theory.[23 25 26 27] Salwen et al.[25] developed the Mott-Smith method by adding an extra term to the two-term Mott-Smith distribution function $f = \sum_{\mu=1}^{3} n_\mu(x) f_\mu$. They could get the correct shock thickness at low Mach numbers, but for the strong shocks other distribution function should be chosen. Radin and Mintzer[26] studied the structure of a strong shock wave using the orthogonal polynomial expansion. Mott-Smith's bimodal distribution function was used as the weighting function to generate the orthogonal polynomials of the expansions. They could not obtain results in cases with the Mach numbers $M < 2.14$.

The Mott-Smith method gives a reasonable agreement with the experimental data and the Monte-Carlo simulation result for strong shocks.[24] As a result, we use the Mott-Smith distribution function in this paper to derive six moments equations for the steady-state problem in one-dimensional domain. By virtue of the system of fluid dynamic equations, the problem of choosing an appropriate velocity moment will be automatically resolved. We will use the collision integral for Maxwell molecules.[28 29] For the case of real particle interaction potential, we will take the temperature dependent viscosity into account.

## FLUID DYNAMICS EQUATIONS

The kinetic equation in a domain of one-dimension takes the following form

$$\frac{\partial f}{\partial t} + V_X \frac{\partial f}{\partial x} = J^B \tag{1}$$

where $f$ is the distribution function of a gas, $t$ the time, and $J^B$ the integral of collisions. In this study we will consider the case of Maxwell molecules.[28 29] The following subset of basic functions is used:

$$\varphi_1 = m, \varphi_2 = mV_X, \varphi_3 = m\xi^2/2,$$
$$\varphi_4 = \frac{m}{2}\xi_X^2, \varphi_5 = \frac{m}{2}\xi_X\xi^2, \varphi_6 = \frac{m}{2}\xi_X^3 \tag{2}$$

In the above, $\vec{\xi} = \vec{V} - \vec{U}$ represents the peculiar velocity and $\vec{U} = (U,0,0)$ is the stream velocity. The same set of basic functions was used in Refs. [30][31] for the problem of wave disturbance propagation in rarefied gas and in Ref. [32] for the description of nonlinear sound propagation in stratified gas following the Grad's method of constructing system of moment equations.[7][8]

Let us define a scalar product in velocity space as follows:

$$<\varphi_I, f> \equiv <\varphi_I> = \int d\vec{V} \varphi_I f$$

The moments of distribution function are related to the thermodynamic variables as follows:

$$\rho = <\varphi_1>, \rho U = <\varphi_2>, \frac{3\rho}{2m}kT = <\varphi_3>,$$
$$P_{XX} = <\varphi_4>, q_X = <\varphi_5>, \bar{q}_X = <\varphi_6>. \tag{3}$$

Here $m$ denotes the mass of a molecule, $k$ the Boltzmann constant, $\rho$ the mass density, $T$ the temperature, $P_{XX}$ the diagonal component of pressure tensor, $q_X$ the vertical component of heat flow, and $\bar{q}_X$ the new parameter having the same unit as the heat flow.

One can project the kinetic equation (1) on the velocity moments in (2) to get the following system of fluid dynamic equations:

$$\frac{\partial}{\partial t}\rho + \frac{\partial}{\partial x}(\rho U) = 0$$

$$\frac{\partial}{\partial t}U + U\frac{\partial}{\partial x}U + \frac{1}{\rho}\frac{\partial}{\partial x}P_{xx} = 0$$

$$\frac{3k}{2m}\frac{\partial}{\partial t}(\rho T) + \frac{3k}{2m}U\frac{\partial}{\partial x}(\rho T) + (\frac{3k}{2m}\rho T + P_{xx})\frac{\partial}{\partial x}U + \frac{\partial}{\partial x}q_x = 0 \quad (4)$$

$$\frac{\partial}{\partial t}P_{xx} + U\frac{\partial}{\partial x}P_{xx} + 3P_{xx}\frac{\partial}{\partial x}U + 2\frac{\partial}{\partial x}q_x = -\frac{p}{\mu}(P_{xx} - \frac{\rho}{m}kT)$$

$$\frac{\partial}{\partial t}q_x + U\frac{\partial}{\partial x}q_x + 2(q_x + \overline{q}_x)\frac{\partial}{\partial x}U - (\frac{3k}{2m}T + \frac{1}{\rho}P_{xx})\frac{\partial}{\partial x}P_{xx} + \frac{\partial}{\partial x}J_1 = -\frac{2}{3}\frac{p}{\mu}q_x$$

$$\frac{\partial}{\partial t}\overline{q}_x + U\frac{\partial}{\partial x}\overline{q}_x + 4\overline{q}_x\frac{\partial}{\partial x}U - \frac{3}{2\rho}P_{xx}\frac{\partial}{\partial x}P_{xx} + \frac{\partial}{\partial x}J_2 = -\frac{p}{\mu}(\frac{3}{2}\overline{q}_x - \frac{1}{2}q_x)$$

where $p$ denotes the pressure and $\mu$ is the viscosity. The above system of equations contains $J_1$ and $J_2$, which are given by

$$J_1 = \int d\vec{V}\xi_x^2 \xi^2 f, \quad J_2 = \int d\vec{V}\xi_x^4 f = <\xi_x^4> \quad (5)$$

To close the above system of equations in Eq. (4), we have to prescribe the distribution function. Several ways can be chosen to close the above moment system. The first is based on the use of polynomial expansions in velocity space for the distribution function. The coefficients in these expansions are, however, the unknown functions of time and space. One may determine them from the moment system, if the number of expansion terms is chosen to be equal to the number of moments.[7] The second method of closing the system of moment equations involves a special choice of the distribution function that is suitable to the specific transport problem under consideration.[11,17,24,30,31,33,34] Simple approximate functions may be chosen if one takes into account the conditions that are specific to the problem. The distribution function is chosen in such a way to get the properties in both of the free-molecular and continuum regimes.

Mott-Smith [11] was the first to use the bimodal Maxwellian distribution function for the description of shock structure. A mixture of two gases of different temperatures, densities and velocities is considered. The Boltzmann equation governs the interaction between these two gases. This idea was then generalized by Lees [33] in an arbitrary curvilinear geometry for the discontinuous distribution function. At present, the distribution functions proposed by Mott-Smith and Lees are considered to be most suitable to solve the solution of boundary value transport problems in a wide range of Knudsen numbers. [18,30,34]

These ideas have resulted successfully in a series of flat and cylindrical (neutral and plasma) Couette flows,[18] in the study of condensation/evaporation of drops of a given size[18] and also in the study of the kinetic Knudsen layer near a cometary nucleus.[35] The approach was developed for the description of nonlinear sound propagation in stratified gas.[32] In Refs. [30,31], the problem of wave disturbance propagation in rarefied gas was studied within the context of the above system. To close the system of differential equations in Eq. (4), a piecewise continuous distribution function[33] was used. The agreement with the experimental data was good for the phase velocity at all Knudsen numbers. The previously proposed moment equations in Refs. [30,31] were derived on the basis of small Mach numbers. Therefore they can not be applied to describe the processes at high Mach numbers.

In this study we will choose the bimodal distribution function,[11] which is proper to describe the subsonic and the other accounting for the supersonic flow:

$$f = f_0 + f_1 \qquad (6)$$

where

$$f_0 = n_0(x)\left(\frac{m}{2\pi kT_0}\right)^{3/2} \exp\left(-\frac{m(\vec{V}-\vec{U}_0)^2}{2kT_0}\right) \tag{7}$$

and similarly for $f_1$ with the subscript 0 being replaced by the subscript 1 throughout.

The parameters $T_0, T_1, \vec{U}_1 = (U_1, 0, 0)$, $\vec{U}_0 = (U_0, 0, 0)$ are assumed to be independent of $x$ and $t$. We'll introduce them in the next section through the Rankine-Hugoniot relations. According to the definitions of density and velocity in (3), we can get the following two expressions through the employed distribution function

$$n_0(x) = n(x)\frac{(U_1 - U(x))}{(U_1 - U_0)},$$

$$n_1(x) = n(x)\frac{(U(x) - U_0)}{(U_1 - U_0)}.$$

The expressions of the integrals $J_{1,2}$ shown in Eq. (5) are given below

$$\begin{aligned}
J_1 &= \frac{n_0(x)}{2}(U-U_0)^4 + 2(U-U_0)^2 n_0(x) V_{T0}^2 + \frac{5}{8} n_0(x) V_{T0}^4 \\
&+ \frac{n_1(x)}{2}(U-U_1)^4 + 2(U-U_1)^2 n_1(x) V_{T1}^2 + \frac{5}{8} n_1(x) V_{T1}^4,
\end{aligned} \tag{8}$$

$$\begin{aligned}
J_2 &= \frac{n_0(x)}{2}(U-U_0)^4 + \frac{3}{2}(U-U_0)^2 n_0(x) V_{T0}^2 + \frac{3}{8} n_0(x) V_{T0}^4 \\
&+ \frac{n_1(x)}{2}(U-U_1)^4 + \frac{3}{2}(U-U_1)^2 n_1(x) V_{T1}^2 + \frac{3}{8} n_1(x) V_{T1}^4
\end{aligned}$$

where $V_T^2 = 2kT/m$.

## SHOCK STRUCTURE

The shock wave, which is stationary in the steady frame of reference under current investigation, connects the equilibrium states of density $\rho_0$, velocity $U_0$ and temperature $T_0$ ahead of the shock at $x \to -\infty$ and the equilibrium quantities $\rho_1, U_1, T_1$ behind the shock at $x \to \infty$. It is convenient to use the dimensionless equations for system (4), where the upstream values are used to define the following dimensionless quantities:

$$\rho' = \frac{\rho}{\rho_0}, U' = \frac{U}{\sqrt{kT_0/m}}, T' = \frac{T}{T_0}, x' = \frac{x}{\lambda_0},$$

$$\pi' = \frac{\pi}{k\rho_0 T_0/m}, q' = \frac{q}{\rho_0 (kT_0/m)^{3/2}}$$

(9)

In the above, $\pi = p_{xx} - \rho kT/m$ and $\lambda_0$ is the mean free path. The mean free path given in Refs. [2,4,36] will be adopted in this study

$$\lambda_0 = \frac{16}{5\sqrt{2\pi}} \frac{\mu_0}{\rho_0 \sqrt{k/mT_0}} \approx \frac{1}{0.783} \frac{\mu_0}{\rho_0 \sqrt{k/mT_0}}$$

(10)

The first three equations, cast in their dimensionless forms (the superscript "prime" in Eq. (9) for the dimensionless variables will be later omitted), in the differential system (4) are as follows:

$$\frac{d}{dx}(\rho U) = 0$$

$$\frac{d}{dx}(\rho U^2 + \rho T + \pi) = 0$$

(11)

$$\frac{d}{dx}(\frac{1}{2}\rho U^3 + \frac{5}{2}\rho TU + \pi U + q) = 0$$

Far ahead of and behind the shock the gas is in equilibrium with $\pi_0 = \pi_1 = 0$ and $q_0 = q_1 = 0$. The dimensionless quantities in front of the shock at $x \to -\infty$ are given by:

$$T_0 = 1, \rho_0 = 1, U_0 = \sqrt{\frac{5}{3}} M_0$$

(12)

Integration of all equations in Eq. (11) between the two equilibrium states gives:

$$\rho_1 = \frac{4M_0^2}{M_0^2+3}$$

$$U_1 = \sqrt{\frac{5}{3}\frac{M_0^2+3}{4M_0}} \quad (13)$$

$$T_1 = \frac{(5M_0^2-1)(M_0^2+3)}{16M_0^2}$$

It is worth noting that use of the above equations, which are well known as the Rankine-Hugoniot relations, enables us to prescribe the boundary conditions.

The number of equations can be reduced further by integrating equations in Eq. (11) from the upstream state to an arbitrary location $x$ in the shock. By taking into account Eq. (12), we get:

$$\rho U = \rho_0 U_0$$
$$\rho U^2 + \rho T + \pi = \rho_0 U_0^2 + \rho_0 T_0 \quad (14)$$
$$\frac{\rho U^3}{2} + \frac{5}{2}\rho T U + \pi U + q = \frac{\rho_0 U_0^3}{2} + \frac{5}{2}\rho_0 T_0 U_0$$

The following relations can be obtained by solving the above three equations in Eq. (14):

$$\rho(U) = \sqrt{\frac{5}{3}}\frac{M_0}{U}$$

$$\pi(U,T) = 1 + \frac{5}{3}M_0^2 - \sqrt{\frac{5}{3}}M_0(\frac{T}{U}+U) \quad (15)$$

$$q(U,T) = \sqrt{\frac{5}{12}}M_0(\frac{5}{3}M_0^2+5+U^2-3T) - U(1+\frac{5}{3}M_0^2)$$

Then we substitute the relations in Eq. (15) into the differential system (4) to get the following system of three coupled ordinary differential equations that govern the transport of velocity $U$, temperature $T$ and $\bar{q}$ below:

$$(3-\frac{4}{3}\sqrt{15}M_0 U + 5M_0^2)\frac{dU}{dx} + 2\frac{d\bar{q}}{dx} = -\frac{p\lambda_0}{\mu\sqrt{kT_0/m}}(1+\frac{5}{3}M_0^2 - \sqrt{\frac{5}{3}}M_0(\frac{T}{U}+U))$$

$$\left(\sqrt{\frac{5}{3}}M_0U^2 + 5\sqrt{\frac{5}{3}}M_0 - \frac{10}{3}UM_0^2 - 2U + 5\frac{\sqrt{15}}{9}M_0^3 - \frac{1}{2}\sqrt{15}M_0T + 2\bar{q}\right)\frac{dU}{dx} + \frac{d}{dx}J_1 - \frac{1}{2}\sqrt{15}M_0U\frac{dT}{dx}$$

$$= -\frac{2}{3}\frac{p\lambda_0}{\mu\sqrt{kT_0/m}}\left(\sqrt{\frac{5}{12}}M_0(\frac{5}{3}M_0^2 + 5 + U^2 - 3T) - U(1 + \frac{5}{3}M_0^2)\right)$$

$$U\frac{d\bar{q}}{dx} + \frac{d}{dx}J_2 + \left(4\bar{q} + \frac{3}{2}U + \frac{5}{2}M_0^2U - \frac{1}{2}\sqrt{15}M_0U^2\right)\frac{dU}{dx}$$

$$= -\frac{p\lambda_0}{\mu\sqrt{kT_0/m}}\left(\frac{3}{2}\bar{q} - \frac{1}{2}\sqrt{\frac{5}{12}}M_0(\frac{5}{3}M_0^2 + 5 + U^2 - 3T) + \frac{1}{2}U(1 + \frac{5}{3}M_0^2)\right)$$

The above three equations can be rewritten in the form given below:

$$A\begin{pmatrix}\frac{d}{dx}U \\ \frac{d}{dx}T \\ \frac{d}{dx}\bar{q}\end{pmatrix} = -\frac{p\lambda_0}{\mu\sqrt{kT_0/m}}\begin{pmatrix}G_1(U,T,\bar{q}) \\ G_2(U,T,\bar{q}) \\ G_3(U,T,\bar{q})\end{pmatrix} \quad (16)$$

where $A$ is the $3*3$ matrix with the nonlinear components. The boundary conditions for the investigated system are specified as

$$T_0 = 1, \; U_0 = \sqrt{\frac{5}{3}}M_0, \bar{q}_0 = 0 \text{ at } x \to -\infty \quad (17)$$

At $x \to \infty$, we impose

$$U_1 = \sqrt{\frac{5}{3}}\frac{M_0^2 + 3}{4M_0}, \; T_1 = \frac{(5M_0^2 - 1)(M_0^2 + 3)}{16M_0^2}, \; \bar{q}_1 = 0 \quad (18)$$

After solving Eq. (16) to get the explicit expressions of three derivatives, we can then solve the coupled first-order ordinary differential equations to get the solutions that connect the fixed points (boundary conditions at $x \to -\infty$ and $x \to \infty$). This is the system for a vector of derivatives $d\mathbf{U}/dx$. By Cramer's rule the system can be solved for the vector of derivatives. There exists, however, a problem if the determinant of $\mathbf{A}$ is zero.[8,9] The critical condition can then be obtained by setting the determinant to be zero. Let's find when the determinant $D$ of $\mathbf{A}(\mathbf{U})$ is zero. At the point $x_0$, the determinant $D_0 = \det\{\mathbf{A}(\mathbf{U}_0)\}$ is zero if $M_0^2 = 1$. At the point $x_1$, the determinant $D_1 = \det\{\mathbf{A}(\mathbf{U}_1)\}$ is zero if $M_0^2 = 1$. Other critical Mach numbers are complex. At $M > 1$ the determinants $D_0$ and $D_1$ are negative. The upper bound of the critical number doesn't exist in our theory at $M > 1$. As a result, in our theory the continuous shock structure exists at all Mach numbers. In the work of Chen[37] there exists an upper critical Mach number $M_C = 2.2$ while in Grad 13 moments method $M_C = 1.65$.[7] This is the reason why they cannot obtain stable solution at Mach numbers larger than $M_C$.

The system of equations was derived on the basis of Gross-Jackson model[21] of Boltzmann equation that corresponds to the special case of Maxwell molecules. The corresponding viscosity is proportional to the temperature following the expression given below with $s = 1$:

$$\mu = \mu_0 \left(\frac{T}{T_0}\right)^s. \tag{19}$$

It is well known that the viscosity takes the same form for other interaction potentials just with an adjustment of the exponent $s$.[5,16,38] For example, $s = 1/2$ is chosen for the hard sphere and $s \approx 0.72$ for the argon.[2,5,16] Other authors[2,5] advised to use $s \approx 0.68$. We will, as a result, use these two values to see which of them can yield a better agreement with the experimental results. According to Eqs. (10) and (19), one gets

$$\frac{p\lambda_0}{\mu\sqrt{kT_0/m}} = \frac{\rho T^{1-s}}{0.783}. \tag{20}$$

In the calculation of Navier-Stokes shock profiles,[1] we will use the constitutive equation to relate the heat conductivity with the viscosity by $\kappa = \frac{15}{4}\mu$ for the case of Maxwell molecules.[31] In the work of Mott-Smith[11] the system of four equations was considered. In this case the density takes the form given below

$$\rho(x) = \rho_1 + \frac{(\rho_0 - \rho_1)}{1 + \exp(\alpha(x/\lambda_0))} \tag{21}$$

where

$$\alpha = \frac{5}{3}\frac{1}{0.783}\sqrt{\frac{3}{5}\frac{1}{M_0}\frac{U_0}{U_1}}\frac{U_0 - U_1}{U_0 + U_1}$$

The profiles of other macroscopic flow quantities may be predicted by the appropriate moments of bimodal distribution function.

## COMPARISON STUDY AND DISCUSSION OF RESULTS

To compute the solutions of temperature and velocity in shock profiles from the proposed system of ordinary differential equations in Eq. (16), subjected to boundary conditions (17) and (18), the computational domain is descretized by $N+2$ positions at $x_i$ with $i = 0, 1, 2..., N+1$ and step size $\Delta x$. The following approximation is used at the nodal point $i$:

$$\left.\frac{dT}{dx}\right|_i = \frac{T_{i+1} - T_{i-1}}{2\Delta x}$$

Calculation of the solutions at positions $x_1$ and $x_N$ requires to know the field values at $x_0$ and $x_{N+1}$, which are given by (17), (18). One needs to derive 3N coupled algebraic equations for the N unknown values of $U$, $T$ and $\bar{q}$. The resulting nonlinear system was solved with the appropriate $\tanh(x)$ curve being considered as an initial guess for the velocity and temperature (similar to Ref. [39]). The predicted temperature and density are presented in a normalized form:

$$\frac{T-T_0}{T_1-T_0}, \frac{n-n_0}{n_1-n_0}$$

One of the main parameters which can well describe shock profile is the shock thickness, which is defined as

$$\delta = \frac{\rho_1 - \rho_0}{\max(\frac{\partial \rho}{\partial x})}$$

The inverse thickness can be derived from Eq. (21) as $\frac{\lambda}{\delta} = \frac{\alpha}{4}$ according to Mott-Smith theory. Another quantity is the temperature-density separation $\Delta_{T\rho}$, which is the distance between two points which have $T = 0.5$ and $\rho = 0.5$, respectively.

In Fig. 1 we compare the results of our work with the results of other authors for the inverse density thickness. For weak shocks, agreement of the solutions between the currently predicted result and the Monte-Carlo simulation result[4] is excellent. Mott-Smith theory[11] predicted a relatively larger thickness at low Mach numbers.

The predicted values of the temperature-density separation $\Delta_{T\rho}$ shown for the Maxwell molecules in Fig. 2 are compared well with the Monte-Carlo simulations,[4] Mott-Smith theory[11] and Navier-Stokes results.[1] Our results are in good agreement with the DSMC calculation in the range of $1 < M < 2.5$. The predicted Navier-Stokes solution is correct only for $M < 1.3$ and the solution calculated by Mott-Smith theory has a good agreement with the DSMC result only in the range of $2.2 < M < 2.5$.

In Figs. 3-10 we compare our results for the density and heat flux profiles with the results of DSMC simulation[40] for Maxwell molecules, Navier-Stokes results and Mott-Smith results computed at different Mach numbers $M = 1.7$, $4$, $8$, $10$. In Figs. 7, 8 we compare our results for the density and heat flux profiles with the results of DSMC simulation of Nanbu[40] and Bird.[16,40] Interrelations between the Nanbu and Bird DSMC methods were shown in Ref.[41]. Nanbu[42] derived his scheme in a mathematical manner directly from the Boltzmann equation. He transformed the Boltzmann equation into a set of equivalent stochastic equations of motion for the simulated molecules. Bird's method[16], which was derived based on the physics of gas flow, is not directly connected with the Boltzmann equation. In Fig. 11 we present the predicted results only for the heat flux profile because it is the higher-order moment of the distribution function and the difference between the underlying theories is clearly seen. The normalized density of our solutions at the coordinate origin $x = 0$ is exactly $0.5$ at any Mach number. Navier-Stokes solutions fail to describe the shock profiles when $M > 1.7$. Our results agree well with the DSMC simulation for both heat flux and density profiles at Mach numbers $1.7 < M < 4$. At $M = 1.7$ and $M = 2$, the heat fluxes predicted from the Mott-Smith theory are larger than the Nanbu DSMC simulation values.[40] At $M = 3$ and $M = 4$, Mott-Smith prediction results for the heat flux lie below the DSMC simulation results. At the large Mach numbers $M = 8$, $10$ our results show a reasonable agreement with the DSMC simulation. In Fig. 7 we can see that at large Mach numbers our theory can well reproduce the DSMC solution of the density profile in the upstream region. In the downstream region Mott-Smith results provide a better agreement with DSMC results. The difference between the predicted results of Nanbu[40] and Bird[16] DSMC simulations can be explained by the statistical errors in DSMC calculations. Our results in the downstream region are close to DSMC results of Bird[16 40]. Small deviation may be caused by the fact that the bimodal distribution function is the approximate solution of the Boltzmann equation. However, we will see from Fig. 15, that for the case of real gas our theory gives excellent agreement with the DSMC simulation results[43] for the density profile in both upstream and downstream parts of the flow. At $M > 4$, the disagreement between the DSMC and Mott-Smith simulation results is apparent for the heat flux profile.

In Fig. 11 we have plotted the results based on the Burnett equations.[36] One can see that the solution curve for the Burnett equations exhibits upstream oscillations. Oscillations appearing in Burnett theory at $M = 1.5$ will increase for the shock investigated at an increasingly higher Mach number.[6 39] Calculation must be carried out very carefully and should be restricted with the step size $\Delta x \approx \lambda_0$. With the decreased step size $\Delta x$, oscillations arise and the convergence of solution cannot be reached. The solution predicted from our proposed equations does not suffer from any oscillation. Application of the super-Burnett equations fails to get rid of the oscillations. As it is shown in Ref.[39], several attempts of improving the Burnett, super-Burnett and Grad equations were reported recently. A good agreement with the experimental and DSMC simulation results was obtained on the basis of Reg13 equations at $M < 4$.[39] However Reg13 equations fail to quantitatively predict strong shock waves.

The temperature profile in Fig. 12 shows its maximum within the shock layer, which can't be predicted by Mott-Smith theory and Navier-Stokes equations. The temperature profile becomes non-monotonic at a Mach number $M > 3$. It is well known that such a predicted temperature profile is not a mathematical artifact but is rather the result of atomistic dynamics.[14 44 45] This overshoot was theoretically predicted firstly by Holway[46]. He showed that the mixture of the two gases of Maxwell molecules can be mixed in such a way that they produce a temperature higher than that of either constituent. The overshoot was later confirmed experimentally (see for example Ref. [13]) as well by the Monte-Carlo[16 24] and molecular dynamics simulations.[15 45] Note that in this article we use six fluid dynamic equations, while Moth-Smith used four equations. In Fig. 12 we have also plotted the results computed from the system of five equations using the same closure procedure. One can see that the increasing number of equations helps to improve the prediction accuracy of shock profile. We expect, as a result, that if we keep increasing the number of moments using the same closure procedure, the agreement with the experiment and Monte-Carlo simulation will be increasingly better.

It is worth noting that the predicted temperature-density separation by Mott-Smith theory is smaller in comparison with the DSMC value. The temperature-density separation by Mott-Smith theory[11] is $\Delta_{T\rho} = 5.89\lambda_0$ at M=10, while in our theory $\Delta_{T\rho}$ is $7.44\lambda_0$, which agrees with the DSMC value.[40] The temperature-density separation predicted by the system of five equations is $\Delta_{T\rho} = 6.50\lambda_0$. For the Navier-Stokes equations we got the value $\Delta_{T\rho} = 3.68\lambda_0$.

In Fig. 13 the predicted values of temperature-density separation are compared with the Monte-Carlo simulation results.[4 40] Our results agree well with the DSMC calculation in the range of Mach numbers $1 < M < 10$. Mott-Smith theory gives good agreement with the DSMC simulation only in the range of Mach numbers $2.2 < M < 2.7$.

In Table I we compare the results of inverse shock thickness predicted by different theories. At $M < 2.5$, the derived system predicts a correct shock thickness. At larger Mach numbers $M > 2.5$, the predicted shock thicknesses differ from the DSMC simulation by an amount 5-15 %. Mott-Smith solution agrees better with the Nanbu solution[40] at large Mach numbers. It is worth noting, however, that agreement of the solutions is only found in the maximum density gradient but not in the density profile itself. In Figs. 3-10 our results are all seen to be in reasonable agreement with the DSMC simulation for Maxwell molecules.

Next, shock parameters are compared with the experimental data and Monte-Carlo simulation for argon. In Fig. 14 the results computed from the derived system are compared with the measured inverse shock thickness for argon.[3] The values $s = 0.72$ and $s = 0.68$ chosen for the viscosity exponent yield a good agreement with the experimental data. We estimate that $s = 0.70$ would provide a result that has the best agreement with the experimental data. Experimental measurement of temperature is difficult and, therefore, Monte-Carlo simulation results are usually used for comparison.

Figs. 15 and 16 show the temperature and density profiles calculated from the DSMC simulation,[43] Navier-Stokes, Brenner-Navier-Stokes[5] and our proposed equations for argon. Argon is modeled with the value $s = 0.72$ for the viscosity exponent in the calculations of the above mentioned theories. Brenner-Navier-Stokes equations produce good agreement in the central and central downstream shock regions, but in the upstream region their predicted profiles are wrong. Our results are in excellent agreement with the DSMC simulation results for both of the density and temperature profiles.

In Fig. 17 we present a higher order moment of the distribution function $\bar{q}_x$ as the function of Mach number. The profiles show the trend similar to heat flux profiles that the peak moves upstream with the increasing Mach number. The magnitudes also increase with the increasing Mach number. In the Mott-Smith theory[11] only the density profile is calculated from the fluid dynamic system, while for the temperature, heat flux and pressure they are calculated from the appropriate moments of bimodal distribution function. Use of the system of six equations instead of four equations in Mott-Smith method helps us to get rid of the incorrect prediction of shock structure at low Mach numbers and gives us some additional insights into the behaviors of temperature, heat flux, pressure in the whole range of Mach numbers. By increasing the number of equations in the proposed formalism, it will allow us to get the information of higher moments of distribution function.

## CONCLUSION

On the basis of Mott-Smith distribution function for the Maxwell molecules, a system of fluid dynamic equations was derived to predict the structure of shock wave in a neutral monatomic gas. The predicted shock thickness is seen to have a good agreement with the Monte-Carlo simulation at all Mach numbers. The predicted inverse density thickness for argon is also shown to be in good agreement with the experimental data. In contrast to the Mott-Smith theory, the derived system predicts the correct shock thickness, temperature-density separation and shock profile at a low Mach number. The Mott-Smith solution is qualitatively correct for $M = 2 \sim 3$. At other Mach numbers their predicted errors are large. Our predicted temperature, density and heat flux profiles are seen to agree well with the Monte-Carlo simulation in the investigated range of Mach numbers $1.7 < M < 10$. We have computed the density and temperature profiles by the present theory and compared with the Monte-Carlo simulation results data for argon. Excellent agreement for density and temperature profiles has been found. In extended thermodynamics, many moments are required in order to get a predicted solution with good agreement with the experimental result. With the Mott-Smith closure, a fairly good agreement with the experimental and the Monte-Carlo simulation results can be obtained even from a differential system with much fewer equations. If one increases the number of moments using the same closure procedure, the agreement with the experimental measurement and the Monte-Carlo simulation is expected to become increasingly better. The proposed procedure can be also applied to the processes involving polyatomic gases, gas mixtures, plasma and problem in astrophysics.

## ACKNOWLEDGEMENT

We would like to thank Manuel Torrilhon for his helpful and constructive discussions. This work was supported by a research grant from the National Science Council of Republic of China with the contract number SC-97-2221-E-002-250-MY3.

**Table 1.** The predicted inverse density thickness for the Maxwell gas using different theories. M-S$_2$, M-S$_3$ – Mott-Smith $V_X^2$ and $V_X^3$ theory,[11] SGZ$_{23}$ - Salwen, Grosch, Ziering ($V_X^3, V_X V^2$),[25] SGZ$_{33}$ - Salwen, Grosch, Ziering ($V_X^3, V_X V^2$),[25] NS- Navier Stokes,[1] MC- Monte-Carlo simulation.[40]

| M | M-S$_2$ | M-S$_3$ | SGZ$_{23}$ | SGZ$_{33}$ | MC | NS | this theory |
|---|---|---|---|---|---|---|---|
| 1.2 | 0.057 | 0.0504 | 0.0653 | 0.065 | | 0.067 | 0.066 |
| 1.5 | 0.124 | 0.116 | 0.136 | 0.143 | | 0.151 | 0.140 |
| 1.7 | 0.154 | | | | 0.164 | 0.194 | 0.167 |
| 2 | 0.184 | 0.193 | 0.192 | 0.212 | 0.193 | 0.242 | 0.193 |
| 2.5 | 0.205 | | 0.200 | 0.224 | 0.202 | 0.286 | 0.201 |
| 3 | 0.206 | 0.251 | 0.196 | 0.223 | 0.205 | 0.300 | 0.196 |
| 4 | 0.188 | 0.248 | 0.170 | 0.193 | 0.186 | 0.288 | 0.168 |
| 5 | 0.165 | 0.228 | 0.146 | 0.165 | 0.163 | 0.262 | 0.145 |
| 6 | 0.146 | | | | 0.145 | 0.235 | 0.127 |
| 7 | 0.129 | | | | 0.128 | 0.211 | 0.112 |
| 8 | 0.115 | | | | 0.116 | 0.189 | 0.101 |
| 9 | 0.104 | | | | 0.105 | 0.172 | 0.090 |
| 10 | 0.0945 | 0.138 | 0.0804 | 0.0902 | 0.0925 | 0.157 | 0.080 |

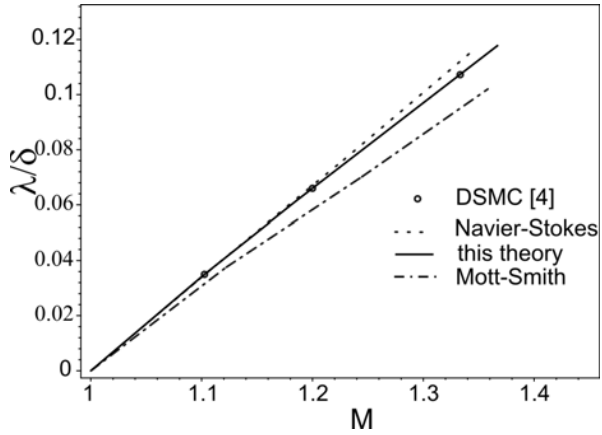

Fig. 1. Comparison of the computed inverse density thicknesses, which are plotted against the Mach number. The currently predicted results are compared with those based on the theories of Navier-Stokes, Mott-Smith, DSMC.

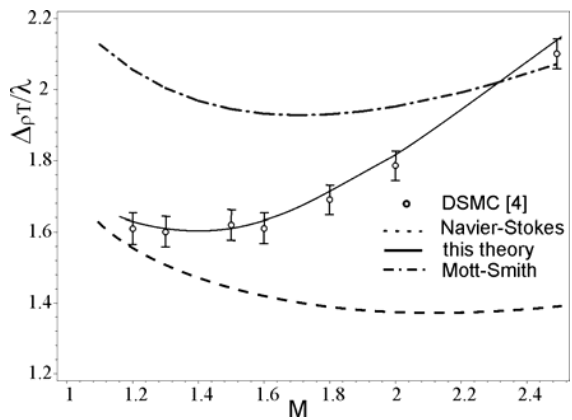

Fig. 2. The plot of the computed values of temperature-density separation which is plotted against the Mach number. Notation – see Fig. 1.

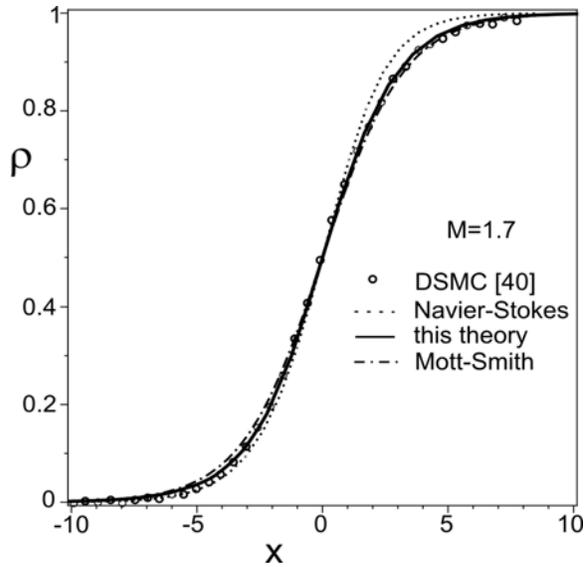

**Fig. 3. Density profile plotted as the function of distance. Comparison of the currently predicted density profile with the DSMC, Navier-Stokes and Mott-Smith simulation results against *x* at *M=1.7*.**

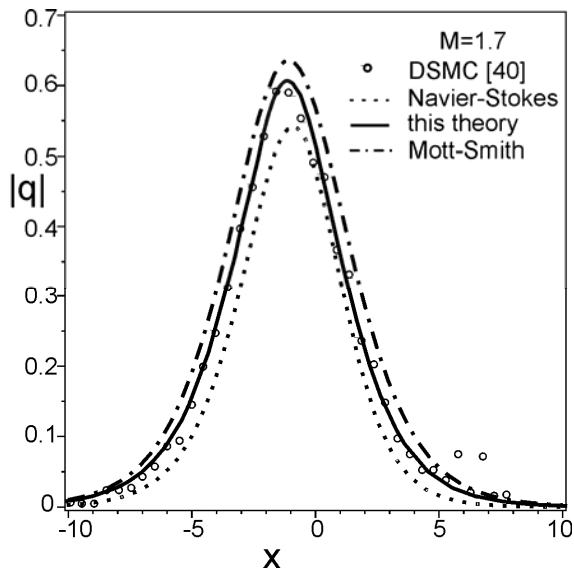

**FIG. 4. Comparison of the heat flux profiles which are plotted as the function of distance. Notation – see Fig. 3.**

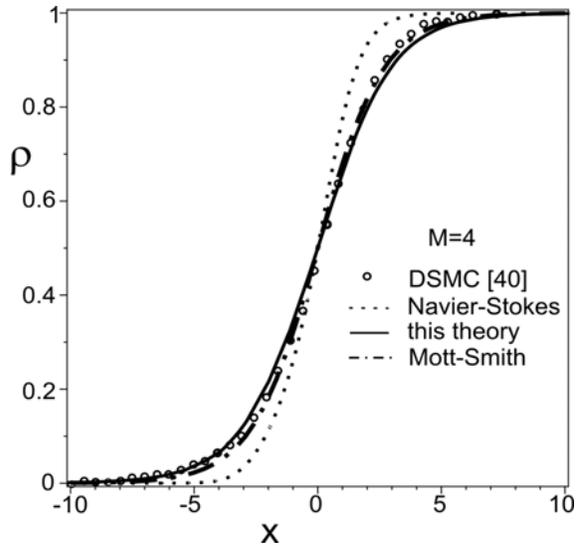

**FIG. 5. The predicted density profiles plotted as the function of distance at M=4.0. Notation – see Fig. 3.**

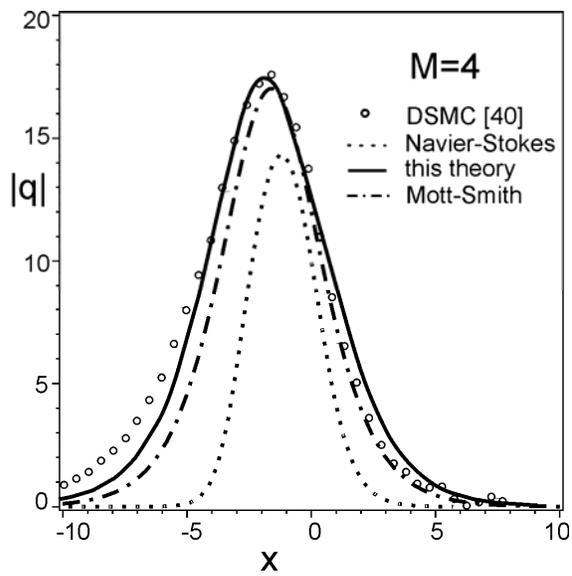

**FIG. 6 – The predicted heat flux profiles plotted as the function of distance at** *M=4.0*. **Notation – see Fig. 3.**

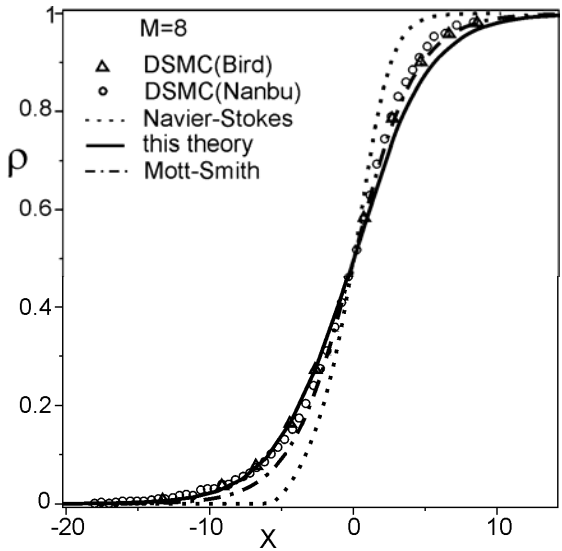

FIG. 7. The predicted density profiles plotted as the function of distance at M=8.0. Comparison of the currently computed results with the results of Nanbu[40] and Bird[16] DSMC, Navier-Stokes and Mott-Smith simulation at *M=8.0*

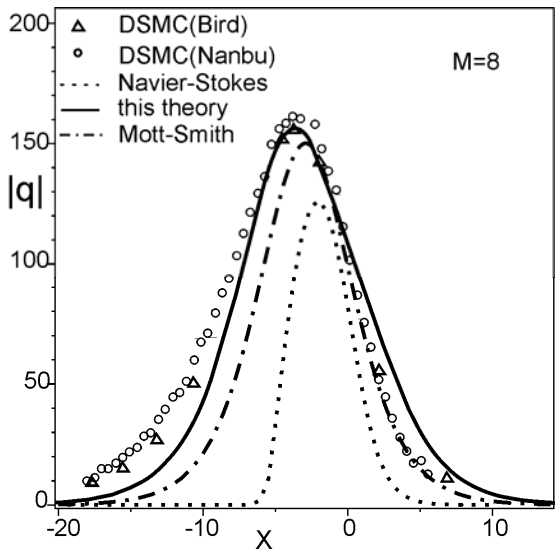

FIG. 8. The predicted heat flux profiles plotted as the function of distance. Comparison of the currently computed results with the results of Nanbu[40] and Bird[16] DSMC, Navier-Stokes and Mott-Smith simulation at *M=8.0*

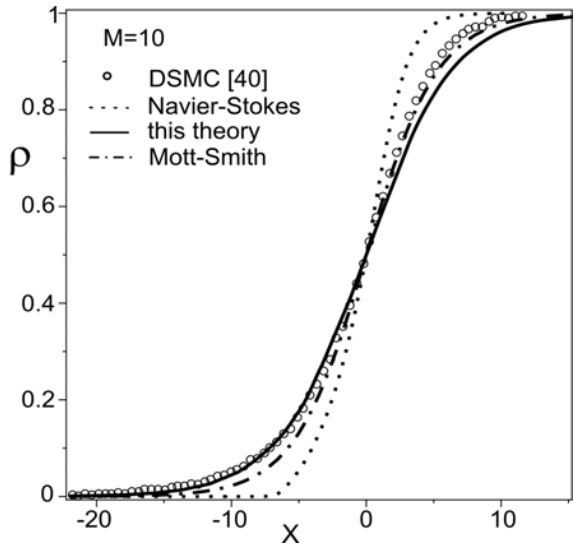

**FIG. 9.** Comparison of the predicted density profiles at *M=10*. Notation – see Fig. 3.

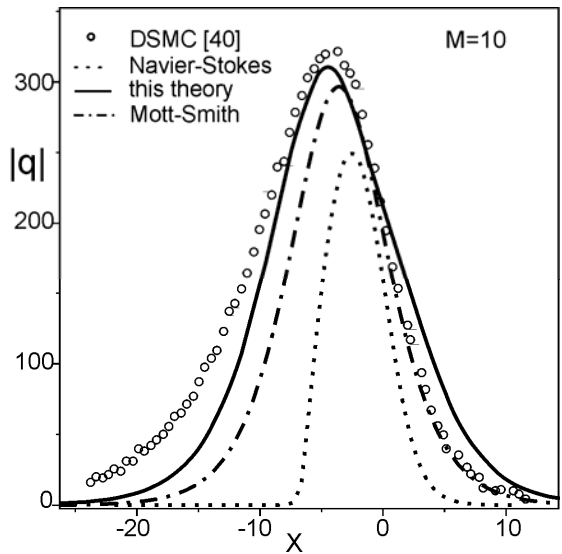

**FIG. 10.** Comparison of the predicted heat flux profiles at *M=10*. Notation – see Fig. 3.

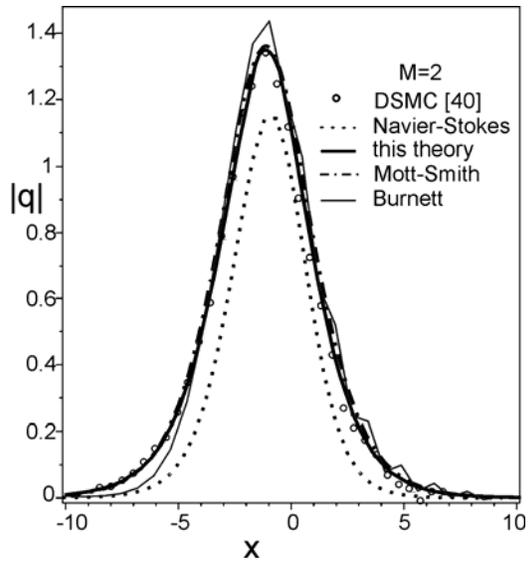

**FIG. 11. Comparison of the predicted heat flux profiles at *M=2*. Notation – see Fig. 3**

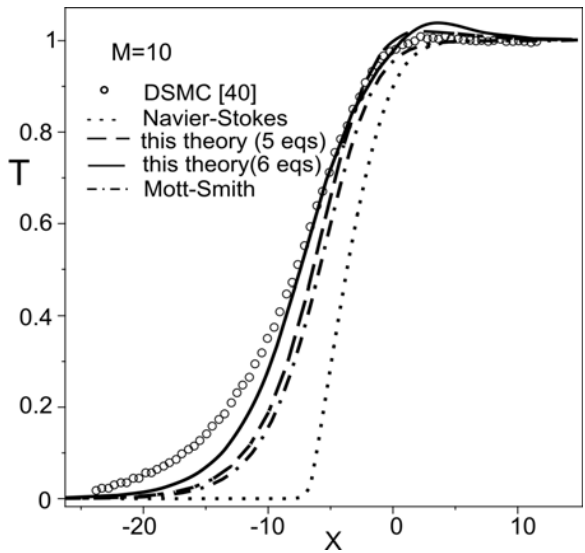

**FIG. 12. Comparison of the predicted temperature profiles at *M=10*. Notation – see Fig. 3. Note that the predicted temperature shows an overshot.**

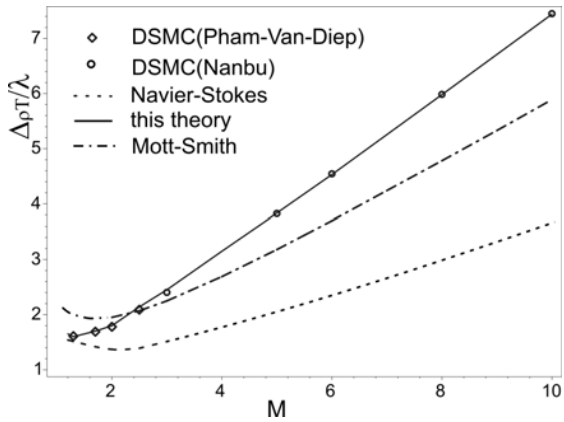

FIG. 13. Comparison of the predicted values of temperature-density separation which is plotted against the Mach number. Diamonds – DSMC results of Pham-Van-Diep[4], circles-DSMC results of Nanbu[40].

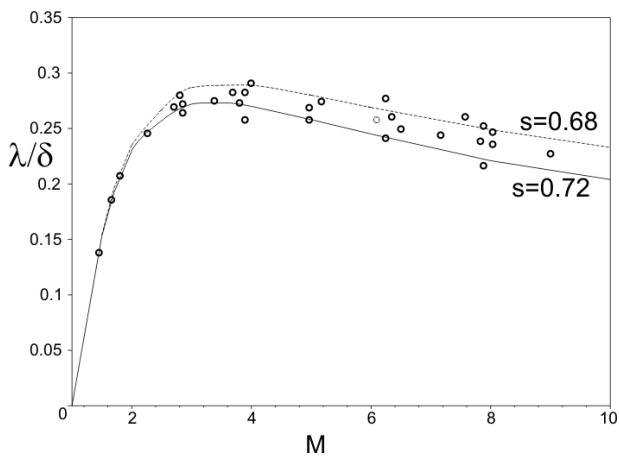

FIG.14 Comparison of the predicted inverse density thicknesses with the experimental data - circles.[2,3]

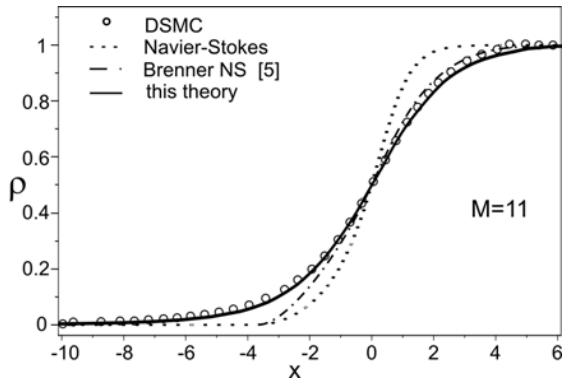

Fig. 15. Density profile plotted as the function of distance. Comparison of the currently predicted density profile with the DSMC, Navier-Stokes and Mott-Smith simulation results against x at M=11; s=0.72.

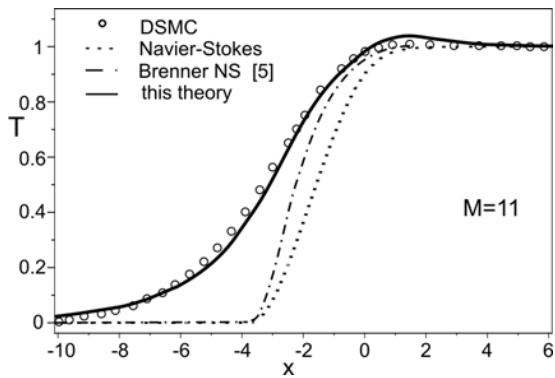

Fig. 16. Temperature profile plotted as the function of distance. Comparison of the currently predicted density profile with the DSMC, Navier-Stokes and Mott-Smith simulation results against x at M=11; s=0.72

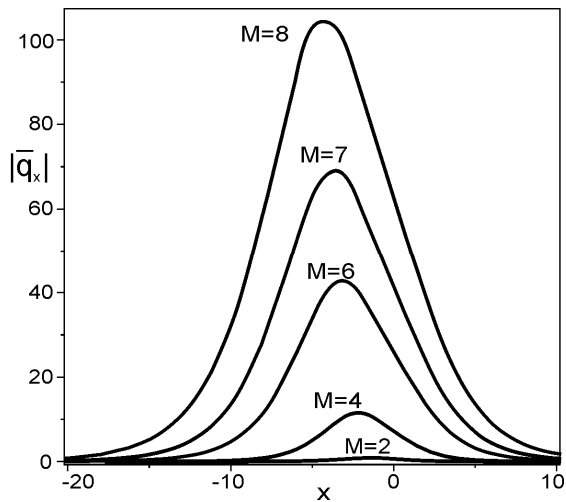

Fig. 17. Plots of the higher order moment of the distribution function $\bar{q}_x$ as the function of distance at different Mach numbers.